\documentclass[usenatbib]{mn2e}
\usepackage[letterpaper,margin=0.8in]{geometry}
\usepackage{hyperref}
\usepackage{natbibmnfix,fixltx2e}
\usepackage{astrojournals}
\usepackage{mathtools}
\usepackage{amsmath}
\usepackage{txfonts}
\usepackage{graphicx}
\usepackage{microtype}
\usepackage{xspace}

\usepackage[utf8]{inputenc}

\usepackage{textcomp}
\usepackage{xcolor}
\usepackage{booktabs,multirow,threeparttable}

\hypersetup{%
  pdftitle={Drag force on G2},
  pdfauthor={\textcopyright\ authors},
  bookmarksopen=true,
  colorlinks=true,
  linkcolor=black,
  citecolor=black,
  urlcolor=black}

\newcommand{\eg}{e.\,g.}

\renewcommand{\deg}{\ensuremath{^{\circ}}\xspace}
\makeatletter

\newcommand{\Rmnum}[1]{\expandafter\@slowromancap\romannumeral #1@}
\makeatother

\newlength{\abovecaptionskip}
\date{}
\pagerange{\pageref{firstpage}--\pageref{lastpage}}
\pubyear{2015}

\renewcommand*{\vec}[1]{\boldsymbol{#1}}

\begin{document}
\label{firstpage}

\title[Drag Force on Gas Clouds]{Going with the flow: using gas clouds to probe the accretion flow feeding Sgr~A$^*$}

\author[McCourt \& Madigan]
{Michael McCourt$^1$\thanks{mmccourt@cfa.harvard.edu}  \& Ann-Marie Madigan$^2$\thanks{Einstein Postdoctoral Fellow; ann-marie@astro.berkeley.edu}
  \vspace{0.1in} 
 \\
$^1$ Institute for Theory and Computation, Harvard University, Center for Astrophysics, 60 Garden St., Cambridge, MA 02138, USA \\
$^2$ Astronomy Department and Theoretical Astrophysics Center, University of California, Berkeley, CA 94720, USA 
}
\maketitle

\begin{abstract}
The massive black hole in our galactic center, Sgr~A$^*$, accretes only a small fraction of the gas available at its Bondi radius.
The physical processes determining this accretion rate remain unknown, partly due to a lack of observational constraints on the gas at distances between $\sim{}10$ and $\sim{}10^{5}$ Schwarzschild radii ($R_{\text{s}}$) from the black hole.
Recent infrared observations identify low-mass gas clouds, G1 and G2, moving on highly eccentric, nearly co-planar orbits through the accretion flow around Sgr~A$^*$. Although it is not yet clear whether these objects contain embedded stars, their extended gaseous envelopes evolve independently as gas clouds. In this paper we attempt to use these gas clouds to constrain the properties of the accretion flow at $\sim{}10^{3}\,R_{\text{s}}$.
Assuming that G1 and G2 follow the same trajectory, we model the small differences in their orbital parameters as evolution resulting from interaction with the background flow.
We find evolution consistent with the G-clouds originating in the clockwise disk.
Our analysis enables the first unique determination of the rotation axis of the accretion flow: we localize the rotation axis to within 20°, finding an orientation consistent with the parsec-scale jet identified in x-ray observations and with the circumnuclear disk, a massive torus of molecular gas $\sim 1.5$ \,pc from Sgr~A$^*$. This suggests that the gas in the accretion flow comes predominantly from the circumnuclear disk, rather than the winds of stars in the young clockwise disk. 
This result will be tested by the Event Horizon Telescope within the next year.
Our model also makes testable predictions for the orbital evolution of G1 and G2, falsifiable on a 5--10 year timescale.
\end{abstract}


\section{Introduction}
\label{intro}
The accretion flows feeding massive black holes have important implications for galaxy evolution, for the growth of massive black holes, for jet dynamics, and for black hole physics in general.
In our galactic center, x-ray measurements determine the density and temperature of the gas at the outer edge of the accretion flow \citep{Baganoff2003,Quataert2002,Quataert2004};
the simplest model for time-steady, non-rotating, adiabatic accretion \citep{Bondi1952} then predicts an inflow rate $\dot{M}_{\text{bondi}}\sim{}10^{-5}\,M_{\odot}/\text{year}$.
However, this predicted accretion rate is inconsistent with almost every other measurement of the gas in the galactic center, including its bolometric luminosity  \citep{Quataert2002}, its rotation measure near the black hole \citep{Agol2000,Quataert2000,Marrone2007}, and the density profile inferred from  x-ray spectroscopy near the Bondi radius \citep{Wang2013}.
Extensions to the Bondi model have therefore been proposed, including convection \citep[e.\,g.][]{Quataert2000}, rotation \citep[e.\,g.][]{Narayan1995,Narayan2011}, magnetic tension \citep{Pen2003,Pang2011}, and outflows \citep[e.\,g.][]{Blandford1999,Quataert2000,Yuan2003}.
These theories lead to different predictions for the density and rotation profiles of the gas in the galactic center; however, a lack of observational probes of the gas at intermediate radii (between $\sim{}10$ and $\sim{}10^5$ Schwarzschild radii, $R_{\text{s}}$) has made it difficult to differentiate these models observationally.

\citet{Gillessen2012} reported the discovery of G2, a gas cloud on an extremely eccentric orbit ($e\sim{}0.98$) about Sgr~A$^{*}$, the massive black hole in the center of our galaxy.
To date, much of the theoretical and observational work on G2 has focused on the possibility that it will disrupt near pericenter ($\sim{}10^3\,R_{\text{s}}$) and drive an episode of increased accretion onto the black hole.
Equally exciting, however, is the opportunity to use measurements of G2 to study the accretion flow feeding Sgr~A$^{*}$: any interaction observed between G2 and the ambient gas it moves through amounts to an important detection of the galactic center accretion flow.
This interaction could take the form of a bow-shock ahead of the cloud \citep{Narayan2012,Sadowski2013b,Sadowski2013a,Yusef2013,Crumley2013}, hydrodynamic disruption of the cloud by shear instabilities \citep{Burkert2012}, or deviation from a Keplerian orbit caused by a (magneto-)hydrodynamic drag force \citep{Pfuhl2015,McCourt2014}.

\citet{Pfuhl2015} present new observations of G1, a gas cloud similar to G2 which was discovered a decade ago \citep{Clenet2005,Ghez2005}.
They find an orbital solution for G1 strikingly similar to that of G2; as we discuss in the following section, this suggests a common origin for the two objects, with G1 preceding G2 by about 13 years.
G1 has a slightly lower semi-major axis and eccentricity than G2, which \citet{Pfuhl2015} explain as the result of a simple drag force due to interaction with the ambient gas.
If true, this represents a detection of the gas in the galactic center, which has so far only been measured at radii much larger, and at radii much smaller, than G2's position.
Using this drag force to infer the properties of the accretion flow at radii of hundreds of AU would provide a valuable test of the various theories for the galactic center accretion flow. 
We explore this possibility here; in particular, we show that the different components of the drag force separately constrain the density and rotation profiles of the accretion flow.

Our results will be especially interesting in coming years as millimeter wave interferometers resolve event horizon-scale structure around the black hole \citep{Doeleman2008,Fish2011}.
Numerical simulations have shown that the gas dynamics very near the black hole is in fact sensitive to the magnetic flux accreted from larger radii \citep[e.\,g.][]{Tchekhovskoy2011,McKinney2012}.
The gas near G2's position thus serves as an outer boundary condition for the horizon-scale accretion disk, and in many ways controls the dynamics that the Event-Horizon Telescope (EHT) will observe.
Moreover, we show in section~\ref{subsec:numeric-results} that the drag force inferred from G2's orbit implies a particular rotation axis for the accretion flow.
Since this rotation axis is not expected to evolve strongly between radii of hundreds of AU and the event horizon,\footnote{unless the black hole spin is misaligned with the accretion disk \citep{McKinney2013}.} this is a prediction that will be tested in the next few years when the EHT images the accretion disk surrounding Sgr~A$^{*}$ \citep{Dexter2010,Broderick2011,Dexter2012,Psaltis2015,Chan2015}.
We discuss this in more detail in section~\ref{sec:discussion}.
Another test will come in early 2018 with the peri-center passage of the S2/S0-2 star which may drive a shock through the accretion flow visible in x-ray emission \citep{Giannios2013}.

A primary uncertainty in using G2 to probe the galactic center accretion flow is that we don't yet know the nature, origin, or structure of the object.
G2 is detected as an extended ($\sim100$\,AU) cloud in re\-com\-bin\-a\-tion-line emission from hydrogen and helium at $\sim\,10^4$\,K and as a $\sim1$\,AU point source in continuum emission, interpreted to be thermally emitting dust at $\sim$\,500\,K.
It is an unusual object and its nature and origin are topics of ongoing investigation
\citep[\eg][]{Burkert2012,Murray2012,Phifer2013,Guillochon2014,DeColle2014};
in particular, it's not yet clear whether the thermal emission represents a star embedded within the cloud \citep[cf.][]{Phifer2013,Witzel2014,Valencia2014}.
Regardless of its origin, however, the ionized component of G2 is spatially extended ($\sim$\,100\,AU) and is measurably distorted by the tidal field of the black hole \citep{Gillessen2012,Gillessen2013a,Gillessen2013b,Pfuhl2015}.
This ionized gas is much larger in extent than the tidal radius ($\sim$\,1\,AU) of any stellar-mass companion, so it behaves independently as an extended cloud.
In this paper, we focus on the dynamics of the ionized gas alone and ignore the central thermal emission.
We thus remain agnostic about the formation scenario for G2.

We have organized this paper as follows: in section~\ref{sec:analytic} we show that the measured orientations of G1 and G2 imply a drag force with a significant component \textit{out} of the orbital plane; such a drag force can only be caused if the background G2 moves through rotates at a significant fraction of the local Keplerian speed.
We present a more detailed analysis in section~\ref{sec:numeric}, where we write down a simplified equation of motion for G2 and numerically integrate its trajectory.
We show that we can reproduce the observed measurements for G1 and G2 with reasonable choices of our model parameters.
We conclude in section~\ref{sec:discussion} by discussing the possibility of using future, improved measurements of G1 and G2 to further constrain the properties of the accretion flow in the galactic center.
We also discuss how our model can be tested using upcoming measurements by the Event Horizon Telescope.

\section{The Torque Connecting G1 and G2}
\label{sec:analytic}
\begin{table}
\centering
\begin{threeparttable}
\caption{(\textit{top}): Angular momenta for G1 and G2 inferred from astrometry and
  velocity data using the MCMC analysis described in appendix~\ref{app:fitting}.
Units of $j$ are $\text{pc}^2/\text{yr}$, and error bars indicate 1-$\sigma$ confidence intervals.
(\textit{bottom}): components of the torque vector derived from the MCMC chain via $\vec{\tau} \sim (\vec{j}_{\text{G1}}-\vec{j}_{\text{G2}})/(t_{\text{peri, G2}}-t_{\text{peri, G1}})$.
We also list the perpendicular and parallel components of the torque, calculated using equations~\ref{eq:tauperp} and~\ref{eq:tauparallel}, respectively.
Torques are normalized to the angular momenta so the units are $[\tau/j]=\text{yr}^{-1}$.
As above, error bars indicate the 1-$\sigma$ confidence interval.}\label{tab:jvectors}
\begin{tabular}
{c%
r@{.}l@{\;±\;}r@{.}l
r@{.}l@{\;±\;}r@{.}l}%
\toprule 
%
{}
& \multicolumn{4}{c}{G2\tnote{1}}
& \multicolumn{4}{c}{G1\tnote{2}} \\
\midrule

$j_x/j$ &     0&14  & 0&05  &    0&36 & 0&07 \\
$j_y/j$ &  $-0$&88  & 0&02  & $-0$&89 & 0&02 \\
$j_z/j$ &     0&46  & 0&03  &    0&29 & 0&02 \\
\midrule

$(j_x)$ &     0&03  & 0&01    &    0&12  & 0&03 \\
$(j_y)$ &  $-0$&20  & 0&02    & $-0$&29  & 0&01  \\
$(j_z)$ &     0&107  & 0&005  &    0&093 & 0&006 \\
\bottomrule
\end{tabular}

\vspace*{\baselineskip}

\begin{tabular}
{c%
r@{.}l@{\;±\;}r@{.}l}
\toprule 
$\langle\tau_x/j\rangle$ &    0&022  & 0&008 \\
$\langle\tau_y/j\rangle$ & $-0$&018  & 0&005 \\
$\langle\tau_z/j\rangle$ & $-0$&007  & 0&003 \\
\midrule

$\langle\tau_{\perp,x}/j\rangle$ &     0&017  & 0&007  \\
$\langle\tau_{\perp,y}/j\rangle$ &  $-0$&001  & 0&003 \\
$\langle\tau_{\perp,z}/j\rangle$ &  $-0$&014  & 0&003 \\
\midrule

$\langle\tau_{\parallel}/j\rangle$ & 0&019  & 0&005 \\

\bottomrule
\end{tabular}
\begin{tablenotes}
\item [1] Gillessen et al. 2013b.

\item [2] Pfuhl et al. 2014.

\end{tablenotes}
\end{threeparttable}
\end{table}


\begin{figure}
  \centering
  \includegraphics[width=\columnwidth]{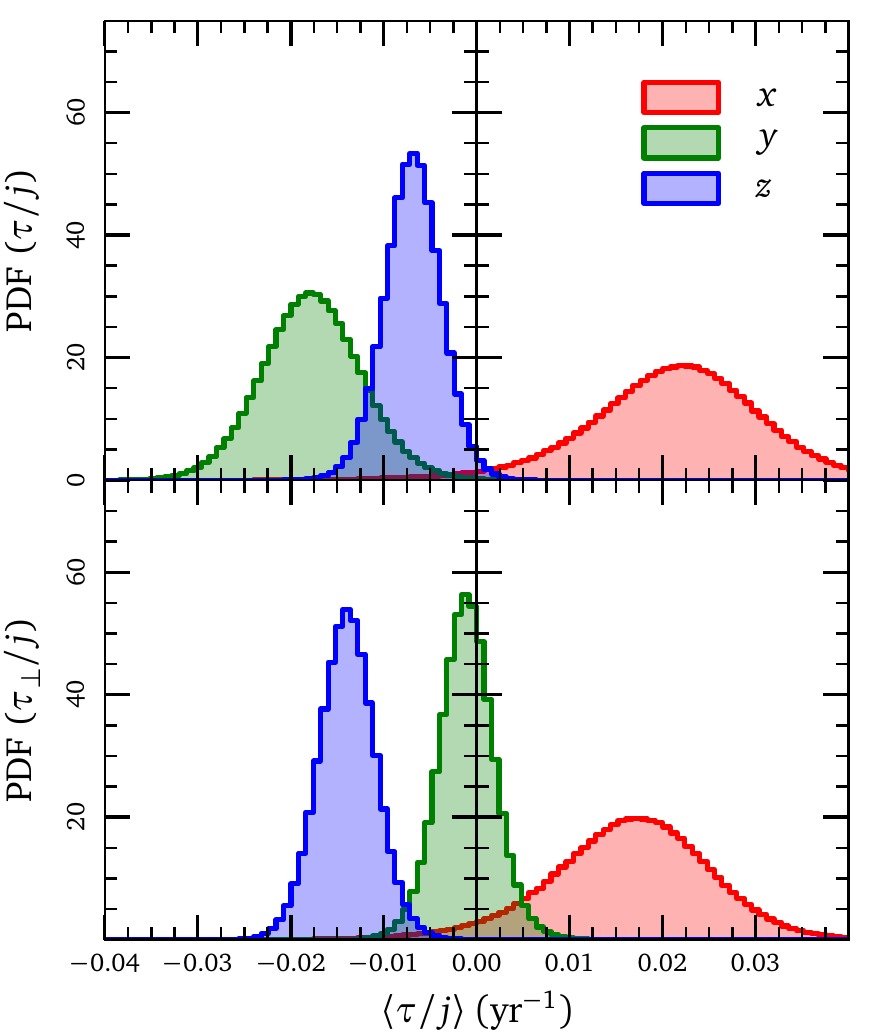}
  \caption{Visualization of the torque vectors presented in table~\ref{tab:jvectors}.
Histograms show the probability distributions for the $x$- (red), $y$- (green), and $z$- (blue) components of the torque vectors computed from the MCMC fit to the data in appendix~\ref{app:fitting}.
The data constrain the perpendicular component of the torque $\vec{\tau}_{\perp}/j$ somewhat more strongly then $\vec{\tau}$ alone, but there is a reasonably strong detection of a non-zero torque in both cases.
This torque is oriented in a way that both re-orients and increases the angular momentum of the cloud; such a torque cannot be produced by a drag force acting on a spherical object moving through a stationary background.
However, we show in section~\ref{sec:numeric} that it can be caused by motion through a rotating accretion flow.}
  \label{fig:tau-pdf}
\end{figure}
\begin{figure}
  \centering
  \includegraphics[width=\columnwidth]{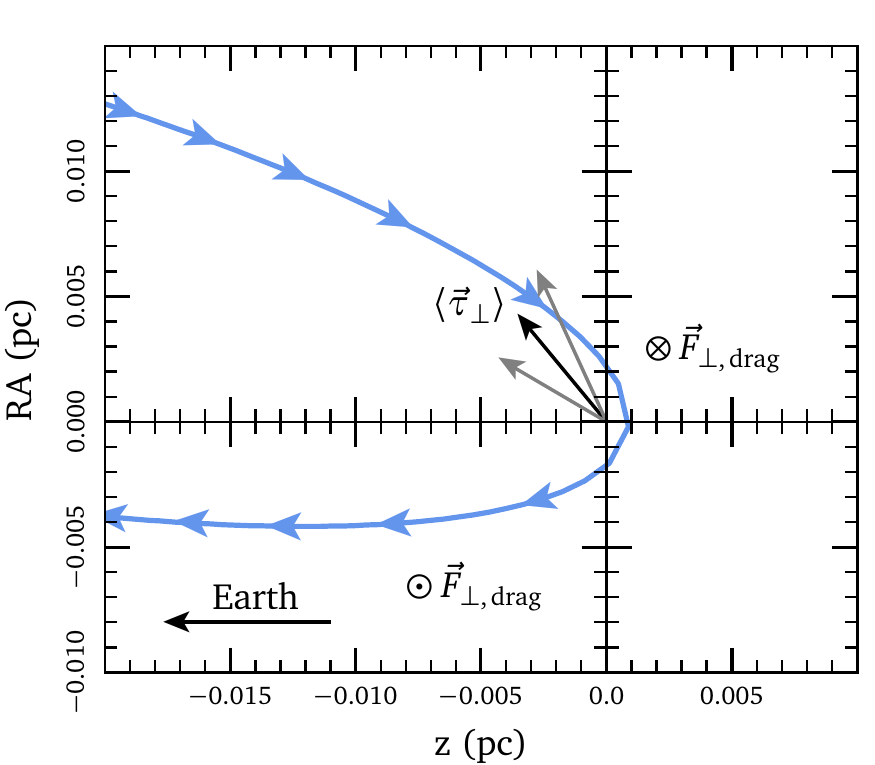}
  \caption{Diagram sketching G2's orbit, along with the perpendicular torque vector required to rotate its orbital plane into G1's orientation.
Gray arrows indicate the 1-$\sigma$ uncertainty in the direction of the torque vector.
This torque implies a force out of the page in the upper-left quadrant of the figure (post-pericenter) and into the page in the lower quadrants (pre-pericenter).
A rotating accretion flow with its rotation axis tilted in the direction of $\vec{\tau}_{\perp}$ can naturally explain this out-of-plane drag force, along with the reversal of the force near pericenter.}
  \label{fig:force-diagram}
\end{figure}
Our analysis is predicated on the assumption that G1 and G2 follow the same trajectory, as suggested by \citet{Pfuhl2015}. 
We begin by discussing why we expect this is a reasonable assumption.
G1 and G2 have strikingly similar properties and are on unusual, highly eccentric orbits \citep{Pfuhl2015}.
The angular momenta of G1 and G2 are aligned to within $(16\pm{}3)$°; if the orbits were isotropically distributed on the sky, there is only a $\sim{}2\%$ probability for such an alignment to occur by chance.
Compounding this probability is the fact that the eccentricity vectors of G1 and G2 are also aligned to within $(29\pm{}4)$°; accounting for the fact that the eccentricity and angular momentum vectors are orthogonal, there is a $\sim{}0.04\%$ chance of drawing two such aligned orbits from an isotropic distribution.
(This probability remains small, $\sim{}1\%$, even if G1 and G2 instead originate from random locations in the clockwise stellar disk.)
Moreover, out of the $\sim{}400$\,year orbital period of G2, the pericenter passages of G1 and G2 coincide to within $\sim{}13$\,years; such temporal alignment has only a $\lesssim{}5\%$ probability of occurring at random.
Together, these probabilities imply that, if G1 and G2 are unrelated objects on independent orbits, we should expect to see $\gtrsim{}10^4$ similar objects within $\sim{}0.04\,\text{pc}$ of Sgr~A$^*$.
Such a population would be readily detectable, even accounting for possible selection biases for clouds with large line-of-sight velocities relative to earth and with epochs of pericenter passage near the present day.

An alternative scenario in which G1 and G2 are related, but somehow following different orbits, also seems improbable.
Since G1 and G2 are on such eccentric orbits, they spend most of their time close to apocenter and likely formed there.
But, if G1 and G2 follow Keplerian orbits, their apocenter locations differ significantly due to their high eccentricities and differing orientations and semi-major axes (see, e.\,g. figure~\ref{fig:3d-orbit-earthcoords}).
Their orbital periods also differ by a factor of $\sim$2--3.
So if G1 and G2 follow separate orbits, they must have formed in very different places and at very different times, only to come into co-location and alignment near their pericenter passages; such a scenario seems contrived.
(We note that the scenario proposed by \citealt{Guillochon2014}, in which G1 and G2 formed near pericenter, does not suffer from this problem; our analysis is consistent with that formation scenario.)

Hence, in this paper we assume that G1 and G2 follow the same trajectory, with G1 preceding G2 by $\sim13$\,years \citep[see][]{Pfuhl2015}. If traveling through a vacuum, this trajectory would be Keplerian. However, the trajectory evolves due to interaction with ambient gas in the accretion flow. The small differences in their orbital parameters thus encode information about the drag force exerted on the trajectory and properties of the accretion flow. 
We model this evolution and attempt to use it to constrain the properties of the accretion flow.

The orientations of the orbital planes for G2 and G1 are known somewhat more precisely than the energies of their orbits \citep{Gillessen2012,Gillessen2013b,Phifer2013,Pfuhl2015}; this constrains the direction of the angular momentum vector better than its magnitude.
Hence, we separately consider two components of the torque vector $\vec{\tau}$:
\begin{subequations}
\begin{alignat}{2}
  \hat{\vec{j}}' 
  &= \left[\boldsymbol{\mathsf{g}} -\hat{\vec{j}}\otimes\hat{\vec{j}} \right]
     \cdot \left(\frac{\vec{\tau}}{j}\right)
  & &= \frac{\vec{\tau}_{\perp}}{j} \label{eq:tauperp} \\
  j' &= \hat{\vec{j}} \cdot \vec{j}' & &= \tau_{\parallel}, \label{eq:tauparallel}
\end{alignat}
\end{subequations}
where $\boldsymbol{\mathsf{g}}$ is the metric tensor (the identity matrix in Cartesian coordinates), $\hat{\vec{j}}$ is a unit vector in the direction of the angular momentum $\vec{j}$, and $j$ is its magnitude.
The perpendicular torque $\vec{\tau}_{\perp}$ reorients the angular momentum $\vec{j}$, while the parallel torque $\tau_{\parallel}$ changes it magnitude.

We estimate the angular momentum vectors for G2 and G1 from the astrometry and velocity measurements published in \citet{Gillessen2013a,Gillessen2013b}\footnote{made publicly available at \url{https://wiki.mpe.mpg.de/gascloud/PlotsNData}} and in \citet{Pfuhl2015}.
We then calculate the components of the torque vector $\tau_{\parallel}$ and $\vec{\tau}_{\perp}/j$ implied by the change in angular momentum.
We provide details of this fit in appendix~\ref{app:fitting}.
Table~\ref{tab:jvectors} lists the angular momentum and torque vectors we obtain.
Figure~\ref{fig:tau-pdf} visualizes this information to illustrate the uncertainty in different components of $\vec{\tau}$ and $\vec{\tau}_{\perp}$.
Though the uncertainties are large, there is good evidence for a change in orbital angular momentum between G2 and G1.

Both the perpendicular component, $\vec{\tau}_{\perp}/j$, and the parallel component, $\tau_{\parallel}/j$, of the torque provide evidence for a rotating accretion flow.
We first discuss the perpendicular component, which re-orients the orbit.
Physically, this torque corresponds to the component of the drag force \textit{out} of G2's orbital plane.
A symmetric object moving through a stationary background cannot experience such a force; our finding that $|\vec{\tau}_{\perp}| \sim \tau$ thus implies either that G2 is flattened like an airplane wing to generate lift, or that it moves through a rotating accretion flow with a rotation velocity comparable to G2's orbital velocity; in fact, we show in section~\ref{subsec:numeric-results} that \textit{both} conditions are necessary to fit the data.
If G2's post-pericenter evolution indeed follows G1, this is a detection of rotation in the galactic center accretion flow.
Figure~\ref{fig:force-diagram} illustrates the geometry of the inferred drag force.

The parallel component of the torque $\tau_{\parallel}/j$, which corresponds to the component of the drag force in G2's orbital plane, also implies a rotating accretion flow in the galactic center.
For the orbital parameters quoted in \citet{Gillessen2013b} and in \citet{Pfuhl2015}, G1 has a \textit{larger} angular momentum than G2.
If a drag force can turn one orbit into the other, it must spin up the orbital angular momentum of the cloud ($\tau_{\parallel}>0$), rather than torquing it down; this is possible only if the background rotates rapidly with a velocity comparable to G2's orbital velocity.
\citet{Pfuhl2015} neglected rotation, and thus could not fit the increase in angular momentum between G2 and G1: they found a larger angular momentum for G2, and smaller energy and angular momentum for G1, than are observed.
If the published orbital elements for G1 and G2 prove accurate, and if G2 continues to follow G1's orbit over the next $\sim10$\,years, this is again a detection of rotation in the galactic center.

The inferred torque vector shown in table~\ref{tab:jvectors} and in figure~\ref{fig:tau-pdf} constitutes the essential result underpinning our analysis.
We quantify it in the next section using numerical integrations and an idealized model for G2's evolution.

\section{Numerical Integrations}
\label{sec:numeric}
\subsection{Method}
\label{subsec:numeric-method}
We model the gas cloud as a point particle moving in the Keplerian potential of the black hole, subject only to a drag force due to its interaction with ambient gas in the accretion flow.
We ignore the stellar contribution to the potential since the timescales for orbital precession and gravitational relaxation are much longer than the orbital period of G2 \citep{Kocsis2011}.

We adopt a drag force $\propto \rho_{\text{bg}} v_{\text{rel}}^2 R_{\text{cloud}}^2 (1+2/[\beta\,M^2])$ as presented in \citet{McCourt2014}, where $\beta \equiv 8\pi{}P_{\text{gas}}/B^2$ quantifies the magnetic field strength in the background plasma, $\rho_{\text{bg}}(\vec{r})$ is the density of the background gas, $\vec{v}_{\text{rel}} \equiv \vec{v} - \vec{v}_{\text{bg}}(\vec{r})$ is the relative velocity between the cloud and the background gas, and $M$ is the Mach number.
Physically, the last term represents an enhancement over the hydrodynamic drag due to the magnetic field in the background plasma.
This leads to the following approximate equation of motion for the cloud:
\begin{align}
  \frac{d^2 r}{d t^2} &= - \frac{G M_{\bullet} \vec{r}}{r^3}
  - \frac{\rho_{\text{bg}} (\vec{r})}{M_{\text{cloud}}}
  \left(1+\frac{2}{\beta M^2}\right) 
  \vec{R}.\label{eq:eom}
\end{align}
The ram pressure vector $\vec{R}$ in equation~\ref{eq:eom} is
\begin{align}
  \vec{R} =
    R_{\text{cloud}}^2 (\vec{v}_{\text{rel}} \cdot \hat{\vec{v}})^2 \hat{\vec{v}} +
    R_{\text{cloud}} L_{\text{cloud}} (\vec{v}_{\text{rel}} \cdot \hat{\vec{a}})^2 \hat{\vec{a}}\nonumber\\ +
    R_{\text{cloud}} L_{\text{cloud}} (\vec{v}_{\text{rel}} \cdot \hat{\vec{b}})^2 \hat{\vec{b}},\label{eq:ram-pressure}
\end{align}
where $\hat{\vec{v}}$, $\hat{\vec{a}}$, and $\hat{\vec{b}}$ form a right-handed coordinate system aligned with the cloud.
We thus allow the cloud to be elongated along its orbit with a (constant) aspect ratio $L_{\text{cloud}}/R_{\text{cloud}}$.
The ram pressure vector $\vec{R}$ reduces to $R_{\text{cloud}}^2{}v_{\text{rel}}^2 \hat{\vec{v}}_{\text{rel}}$ when $\vec{v}_{\text{rel}}$ is parallel to $\vec{v}$.
If the cloud is quasi-spherical, with $L_{\text{cloud}} = R_{\text{cloud}}$, the drag force points approximately in the direction opposite the relative velocity $\vec{v}_{\text{rel}}$.\footnote{not precisely opposite, because equation~\ref{eq:ram-pressure} is appropriate for a cloud with flat surfaces, rather than a sphere. Due to the lack of perfect symmetry, non-spherical clouds can experience lift.}
A non-spherical cloud develops a drag force in a direction different from $\vec{v}_{\text{rel}}$; this force is akin to aerodynamic lift and we show in section~\ref{subsec:numeric-results} that it is essential to fit the data.

Throughout our analysis, we adopt the best-fit values for the mass of the black hole, $M_{\bullet} = 4.3 \times 10^6\,M_{\odot}$, and for the distance to the galactic center, $R_0 = 8.3\,\text{kpc}$, from \citet{Gillessen2009}.
We also take $M_{\text{cloud}} \sim 3\,M_{\text{earth}}$ and $R_{\text{cloud}} \sim 100\,\text{AU}$, as suggested in \citet{Gillessen2012}.
We do not vary these parameters when attempting to fit our model to the data.

In order to derive an initial position and velocity, we took the astrometry and velocity data for G2 from \citet{Gillessen2013a,Gillessen2013b}.
We fit the observed positions and line-of-sight velocities with second-order polynomials in time and use these fits to produce an initial condition for G2 at 2013.33.
In units of parsecs and years, this reads:
\begin{subequations}
\begin{align}
\vec{r}\,(2013.33) &= \left[
  \begin{array}{r@{.}l}
   0&003537\\
  -0&00002680\\
  -0&001088\\
  \end{array}
  \right] \\
\vec{v}\,(2013.33) &= \left[
    \begin{array}{r@{.}l}
    -0&002075\\
     0&0008509\\
     0&002189
    \end{array}
    \right],
\end{align}
\end{subequations}
where the origin is centered on Sgr~A$^*$ and the $x$-, $y$-, and $z$-axes correspond to right ascension, declination, and the line-of-sight distance, respectively.
This initial condition is consistent with the published Kepler elements for G2's orbit \citep{Gillessen2012,Phifer2013}.
Starting from this initial condition, we integrate the equations of motion using the \texttt{NDSolve} function in Mathematica.
We have made our code publicly available on github.\footnote{\url{https://github.com/mkmcc/g2-drag-force}}

The background accretion flow enters into our equation of motion through the density $\rho_{\text{bg}}(\vec{r})$, the relative velocity $\vec{v}_{\text{rel}} \equiv \vec{v} - \vec{v}_{\text{bg}}(\vec{r})$, the magnetic field strength $\beta$, and the Mach number $M$, which depends on the temperature of the background gas.
Unfortunately, there are very few observational constraints on the background accretion flow at the range of radii where G2 orbits.
We proceed by adopting a model for the background gas; we then compare resulting trajectories with measurements of G1 and G2 to constrain the parameters in our model.
In this paper, we adopt an extremely simplistic model for the density, temperature, and velocity of the background gas:
\begin{subequations}
\begin{align}
  \rho_{\text{bg}}(\vec{r}) &= 
    \rho_0 \left(\frac{r}{r_0}\right)^{-\alpha} \\
  T_{\text{bg}}(\vec{r}) &= \frac{G M_{\bullet}}{r} \\
  \vec{v}_{\text{bg}} (\vec{r}) &= 
    f_{\text{kep}} \left(\frac{G M_{\bullet}}{r}\right)^{1/2}
    \frac{\vec{J}\times\vec{r}}{J\;r},
\end{align}\label{eq:background}
\end{subequations}
where $\rho_0 = 1.7\times{}10^5\,M_\text{Earth}/\text{pc}^3$ and $r_0 = 0.04\,\text{pc}$ are fixed by x-ray measurements of Bremsstrahlung emission near the Bondi radius \citep{Quataert2002,Quataert2004}.
This model is specified by the power-law exponent $\alpha$, by the rotation parameter $f_{\text{kep}} = v_{\text{bg}}/v_{\text{circ}}$, by the two angular coordinates ($\theta, \phi$) of the angular momentum vector $\vec{J} \propto [\sin(\theta)\cos(\phi), \sin(\theta)\sin(\phi), \cos(\theta)]$, and additionally by the assumed magnetic field strength (via $\beta$ in equation~\ref{eq:eom}).
This model does not represent an internally consistent theory for the accretion flow in the galactic center \citep[see, \eg,][]{Yuan2003}; rather, it is a minimal model which might rotate the orbital plane of an infalling gas cloud.
It is straightforward to extend our analysis by replacing equation~\ref{eq:background} with a suitable model for the gas in the galactic center; the analysis discussed in the next section might then be used to differentiate among the various theories for the accretion flow.
Due to the limited amount of data for G1 and G2 and the uncertainty in their masses and shapes, we cannot yet do so. However, we describe in section~\ref{sec:discussion} how continued observations of G1 and G2 could improve our constraints on a timescale of $\sim$5--10\,years.

After integrating each trajectory, we estimate its likelihood assuming the probability $P \propto \exp(-\chi^2)$, where 
\begin{align}
  \chi^2 = \sum_{i} \frac{[\text{model}(t_i)-\text{data}_i]^2}{\text{err}^2_i}\label{eq:chi-sq-def}
\end{align}
is calculated using the published astrometry and velocity measurements for G1 \citep{Pfuhl2015} and G2 \citep{Gillessen2013a,Gillessen2013b}.
We maximize this likelihood over all initial conditions and background models. We use two complimentary methods for this optimization. First, we use Mathematica's built-in simulated annealing optimizer, which produces a population of orbits tightly clustered around the best-fitting solution; we show this best-fitting orbit in figures~\ref{fig:2d-orbit}, \ref{fig:3d-orbit}, and~\ref{fig:kepler-elements}.
Second, we use the \texttt{emcee} MCMC optimizer \citep{Foreman2013}, which is an implementation of the affine-invariant algorithm presented in \citet{Goodman2010}.
We use the resulting Markov chain to estimate the uncertainty in our model parameters (figures~\ref{fig:scatter-plots} and \ref{fig:kepler-elements}); however we caution that the uncertainty in our results is likely dominated by simplifications in our model, not by the statistical noise quantified by the Markov chain.
By analogy with the Gaussian distribution, we define the ``1-$\sigma$'' uncertainty for the parameters in our Markov chain as the interval between the $d$ and $1-d$ quantiles of the data, where $d = \frac{1}{2}\text{erfc}\left(1/\sqrt{2}\right) \sim 0.159$.
For the ``2-$\sigma$'' uncertainty interval, $d = \frac{1}{2}\text{erfc}\left(\sqrt{2}\right) \sim 0.0228$.
The best-fit values we quote for the Markov chain are the most probable values determined via the binning method described by \citet{Knuth2006}.

We find the best-fitting orbit has $\chi^2/\text{d.o.f.}\sim4$, rather than a value $\sim{}1$ expected for a good fit to the data.
As discussed in appendix~\ref{app:fitting} and in the following section, this is comparable to the $\chi^2/\text{d.o.f.}$ we obtain by fitting G1 and G2 to separate Keplerian orbits; it thus seems likely that the errors quoted for the observational measurements are underestimated by a factor of $\lesssim2$.
Following \citet{Guillochon2014}, we quantify this additional error by repeating our analysis with the modified statistic:
\begin{align}
  P \propto \prod_i \frac{1}{\sqrt{2\pi s_i^2} }
  \exp\left(
    - \frac{[\text{model}(t_i)-\text{data}_i]^2}
           {s_i^2}
   \right),\label{eq:mla-probability}
\end{align}
where
\begin{align}
  s_i^2 = \sigma_i^2 + f^2 \langle \text{data} \rangle^2,\label{eq:mla-sigma}
\end{align}
and the extra parameter $f$ represents a possible systematic uncertainty in addition to the error-bar $\sigma$ quoted for the observational points.
We allow different $f$'s for astrometry and line-of-sight velocity measurements.
This statistic produces similar results to the $\chi^2$ statistic mentioned above, but yields a more reasonable likelihood with an equivalent $\chi^2/\text{d.o.f.}\sim1.5$.

In our optimization, we assume flat priors on $\alpha$, $f_{\text{kep}}$, and on perturbations to the initial condition.
We assume the Jeffreys prior for the parameter $f$.
The polar and azimuthal angles $\theta$ and $\phi$ define the orientation of the rotation axis; we assume these are isotropically distributed, i.\,e. with flat priors for $\sin(\theta)$ and for $\phi$.
We also assume flat priors for $\log\,\beta$ and for $\log\,(L_{\text{cloud}}/R_{\text{cloud}})$.
While these are not necessarily ``uninformative'' priors, different choices do not appear to bias our results by more than $\sim$\,1-$\sigma$.
Different choices for the background model in equation~\ref{eq:background} have a larger effect.

Equations~\ref{eq:eom}--\ref{eq:background} fully specify our numerical model for G2's evolution, and equations~\ref{eq:chi-sq-def}--\ref{eq:mla-sigma} specify our comparison to the data.
We show our results in the following section.

\subsection{Results}
\label{subsec:numeric-results}
\begin{figure*}
  \centering
  \includegraphics[width=\textwidth]{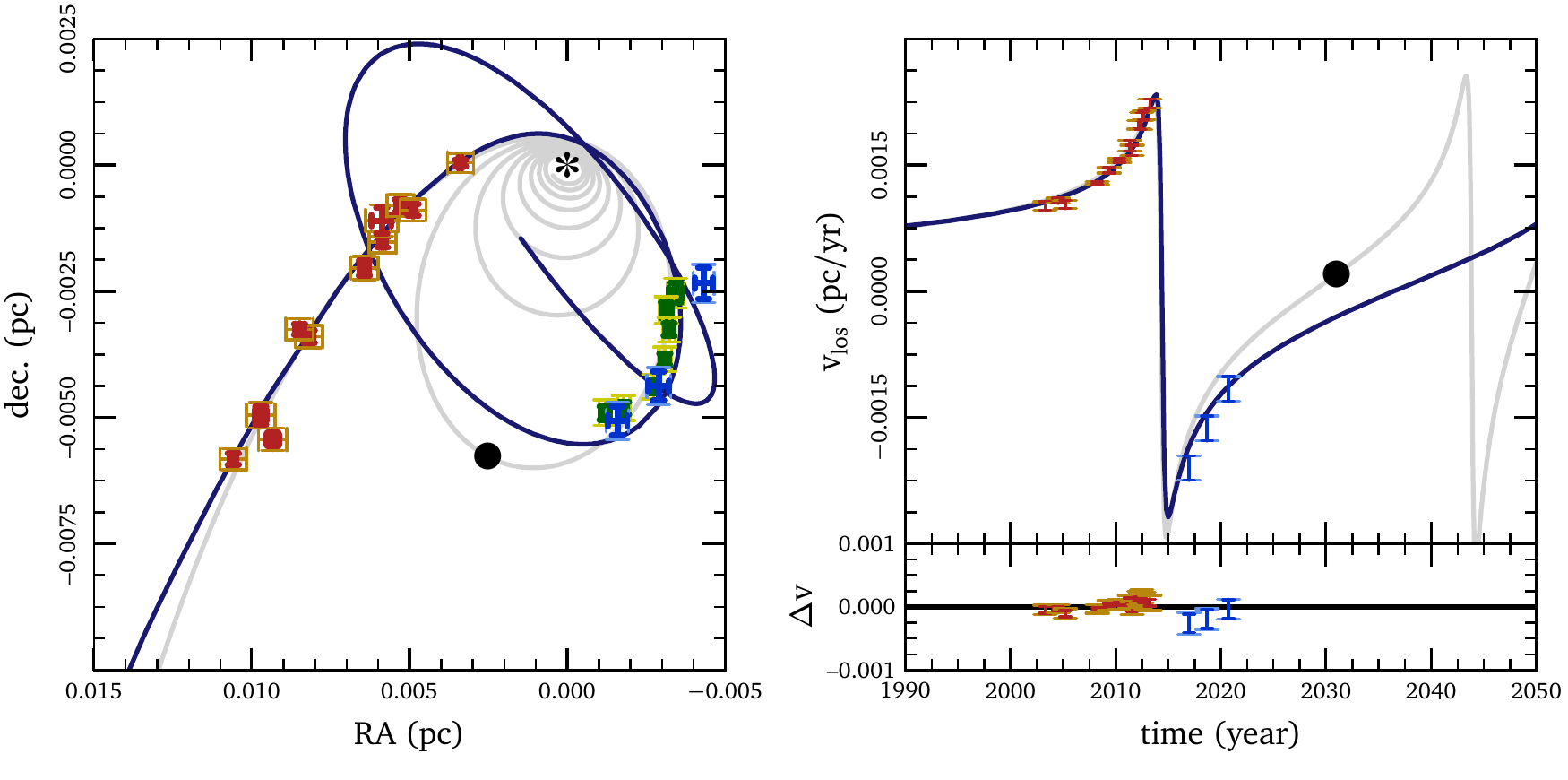}
  \caption{(\textit{left:}) Comparison of best-fit model (blue line) with astrometric data.
Smaller, dark error bars correspond to those reported in \citet{Gillessen2013a,Gillessen2013b} and  \citet{Pfuhl2015}.
Larger error bars include the systematic error found from our maximum-likelihood analysis (see section~\ref{subsec:numeric-method}).
Though the cloud appears to be swept away in this figure, this is a projection effect; the orbit does shrink and circularize (see figure~\ref{fig:3d-orbit}).
The light gray orbit shows a model in which the drag force is restricted to the plane of the orbit, as presented in \citet{Pfuhl2015}.
The location of Sgr~A$^{*}$ is marked with an asterisk. 
(\textit{right:}) Comparison of best-fit model with line-of-sight velocity.
As in the astrometry plot, the smaller error bars are from \citet{Gillessen2013a,Gillessen2013b} and \citet{Pfuhl2015}, while the larger ones account for the systematic error determined by our maximum-likelihood analysis. In both panels, the filled black circles mark the location of G1 in 2018 around the time of the peri-center passage of S2/S0-2, for the \citet{Pfuhl2015} model. The differences with our model (blue line) will be easily detectable.
}\label{fig:2d-orbit}
\end{figure*}
\begin{figure}
  \includegraphics[width=\columnwidth]{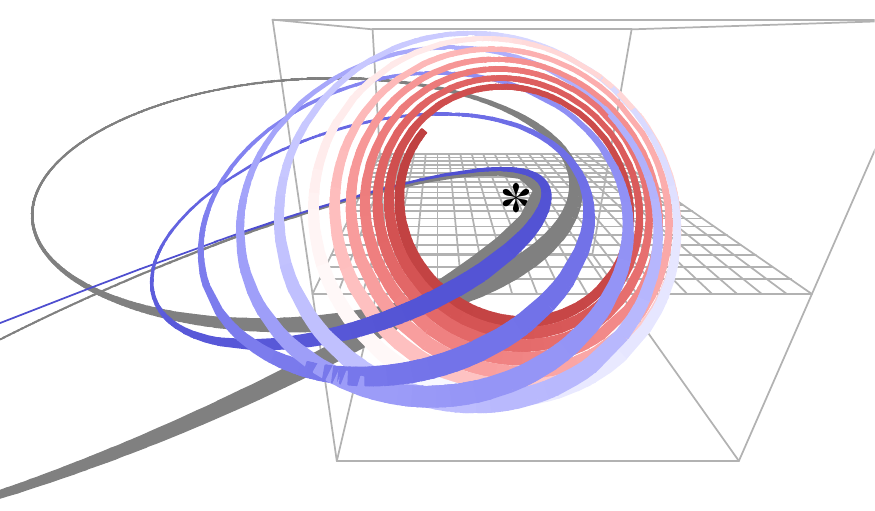}
  \caption{Perspective rendering of the best-fit orbit.
Gray ellipses show the Keplerian fits for G2 and for G1.
In our model, the gas cloud initially comes in from apocenter approximately following the Keplerian orbit fit to G2.
The cloud's orbital plane is misaligned with the rotation of the background flow, resulting in a perpendicular component to the drag force.
The resulting torque rapidly re-orients the orbital plane and,
after $\sim$\,5 pericenter passages, the orbital plane is aligned with the rotation axis of the accretion flow.
No further reorientation occurs, but the orbit circularizes and the cloud becomes approximately co-moving with the rotating gas.
The drag force thus drops significantly; the cloud slowly sinks in toward the black hole.
Figure~\ref{fig:3d-orbit} shows this orbit from several different angles to better illustrate the rotation.}
\label{fig:3d-orbit-earthcoords}
\end{figure}
\begin{figure*}
  \includegraphics[width=\textwidth]{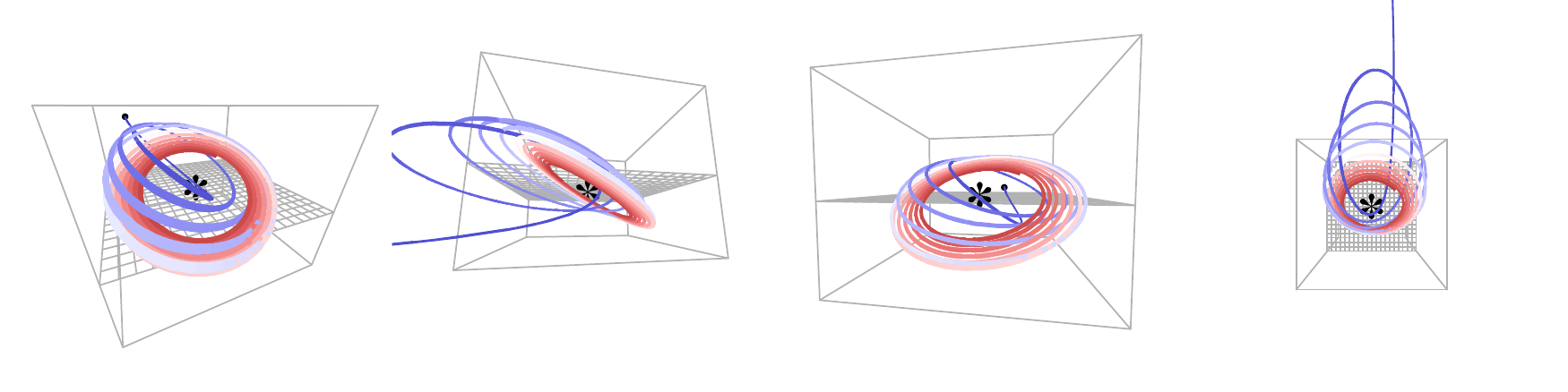}
  \caption{Perspective rendering of the best-fit orbit.
In order to more clearly show the reorientation of the orbit, the coordinate system is rotated so that apocenter lies along the $x$-axis and the rotation axis of the accretion flow lies in the $x$-$z$~plane.
The cloud initially plunges toward the black hole, and the drag force quickly re-orients the orbital plane after $\sim$5 pericenter passages.
Once the orbit aligns with the accretion flow, it becomes approximately co-moving with the gas.
The drag force is thus much reduced, and the orbit circularizes and slowly moves in from that radius.}\label{fig:3d-orbit}
\end{figure*}
\begin{figure*}
  \centering
  \includegraphics[width=\textwidth]{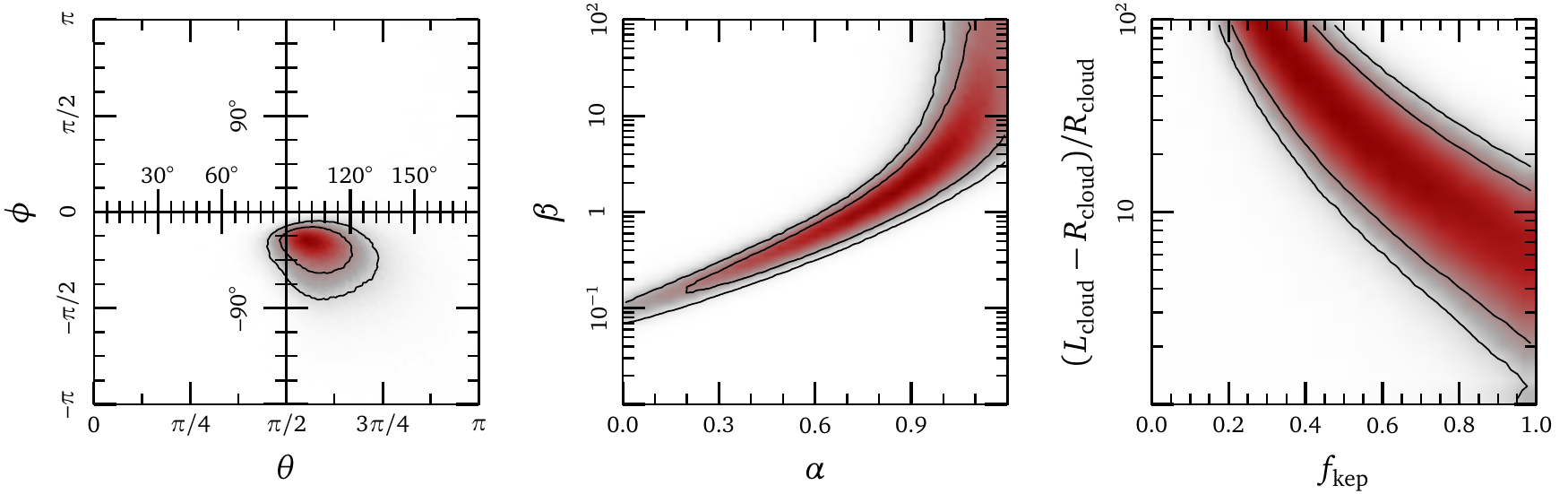}
  \caption{Probability distributions of the parameters in our model, as determined by our maximum-likelihood analysis (we obtained very similar results from the $\chi^2$ analysis described in section~\ref{subsec:numeric-method}).
    The data constrain some parameters much more strongly than others: for example, we localize the orientation of the rotation axis of the accretion flow  to within $\sim\,20\deg$ (\textit{left panel}).
    The parameters $\alpha$ and $\beta$ both control the magnitude of the drag force; we thus find they are strongly degenerate (\textit{center panel}).
    The magnetic field strength can be determined by other observations, however \citep[e.\,g.][]{Eatough2013}; this would enable a constraint on the gas density in the galactic center.
    Similarly, the rotation parameter $f_{\text{kep}}$ and the cloud aspect ratio $L_{\text{cloud}}/R_{\text{cloud}}$ control the component of the drag force out of the plane; these parameters are also strongly degenerate (\textit{right panel}).
    The shape of the cloud can be inferred from modelling its velocity shear as determined by Br-$\gamma$ observations; this would enable a constraint on the rotation profile of the accretion flow.}
\label{fig:scatter-plots}
\end{figure*}
\begin{figure}
  \centering
  \includegraphics[width=\columnwidth]{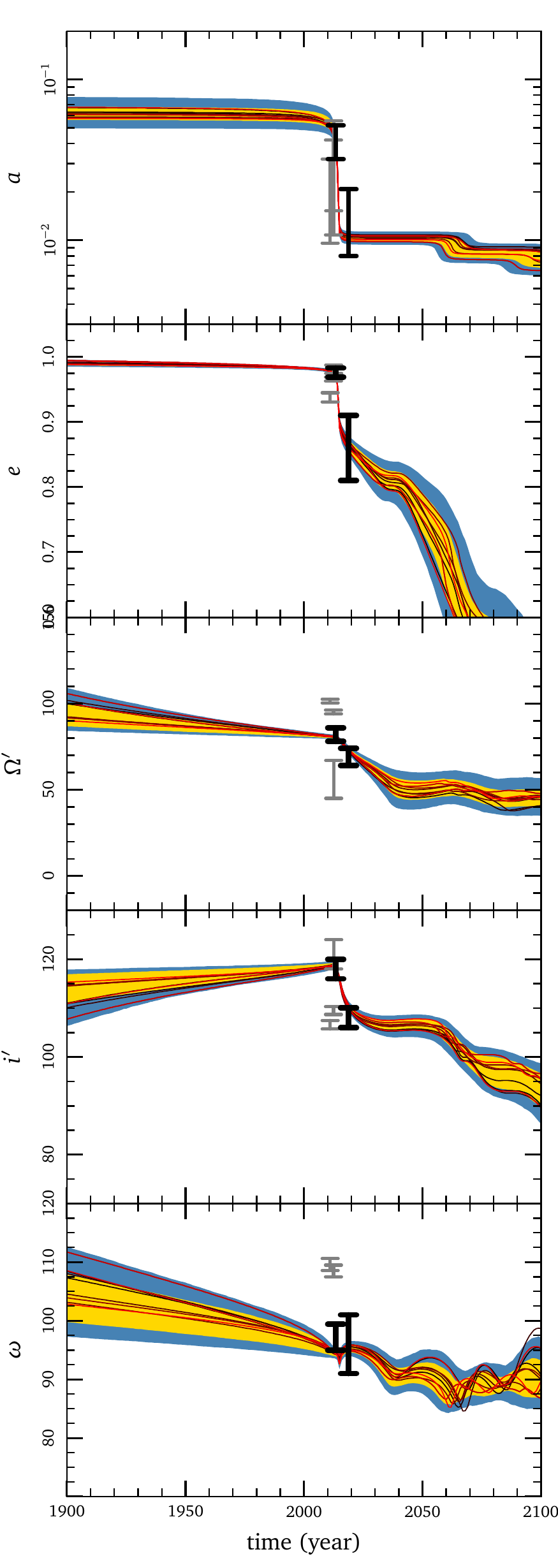}
  \caption{\textit{caption on next column.}}
  \label{fig:kepler-elements}
\end{figure}
\begin{figure}
  \contcaption{Comparison of the evolution of Kepler elements with time for some well-fitting orbits with derived values for G1 and G2. Red lines show a random sample of well-fitting solutions. Yellow and blue bands indicate 1- and 2-$\sigma$ intervals. Black points are from \citet{Gillessen2013b} and from \citet{Pfuhl2015}; we use the data from these papers in our analysis.
  For comparison, the gray points show results from \citet{Gillessen2012}, \citet{Gillessen2013a}, and \citet{Phifer2013}.}
\end{figure}
The dark blue curve in figure~\ref{fig:2d-orbit} shows the best-fitting orbit we found with the drag force model in equation~\ref{eq:eom}.
In this figure, the red points indicate the Br-$\gamma$ observations of G2 from \citet{Gillessen2013b} and the green and blue points respectively show the L-band and Br-$\gamma$ observations of G1 from \citet{Pfuhl2015}.
This solution has a $\chi^2/\text{d.o.f.}=\{4.2, 1.4\}$, where the first number is a standard $\chi^2$ defined using equation~\ref{eq:chi-sq-def} and the second number is calculated using the larger error bars found via equation~\ref{eq:mla-probability}.
(We show these larger error bars in light colors in figure~\ref{fig:2d-orbit}.)
This quality of fit is essentially equivalent to the best fit we obtained while fitting G1 and G2 to separate Kepler orbits ($\chi^2/\text{d.o.f.}=\{4.4, 1.5\}$; see appendix~\ref{app:fitting}); the data thus do not presently favor our model over one in which G1 and G2 are unrelated objects on distinct Keplerian orbits.
However, as discussed in section~\ref{sec:analytic}, G1 and G2 are on unusual orbits and have very different apocenter locations; if they are unrelated objects, they formed in different places and at different times, and it thus seems highly improbable to find them so closely co-located at pericenter.
This low probability is not accounted for in the $\chi^2$ statistic quoted above, but should be considered when evaluating a model in which G1 and G2 are unrelated.

The gray curve in figure~\ref{fig:2d-orbit} shows a model in which the drag force is restricted to the orbital plane, as in \citet{Pfuhl2015}.
This model fits the data similarly to ours ($\chi^2/\text{d.o.f.} = \{5.6, 1.7\}$), but allowing for rotation more closely matches the post-pericenter radial velocity.
While these two models both fit the existing data, they predict very different evolution following the first pericenter passage.
For G1, which is ahead of G2 by about a decade, the two models will begin to diverge appreciably in 5--10\,years.

Figure~\ref{fig:3d-orbit-earthcoords} shows a perspective rendering of our best-fit orbit, along with the Kepler ellipses fit to G1 and G2 by \citet{Pfuhl2015} and \citet{Gillessen2013b}.
While the cloud's trajectory does lie tangent to the Kepler ellipses for G1 and G2, this alignment is transitory.
This figure is drawn in a coordinate system with declination increasing ``up'' the page, line-of-sight distance increasing to the right, and right ascension increasing into the page. 
Figure~\ref{fig:3d-orbit} shows the same orbit in a rotated coordinate system aligned such that the cloud is initially on the $x$-axis, and the rotation axis of the background flow lies in the $x$-$z$~plane.
This coordinate system more clearly shows the rotation of the cloud's orbital plane.
In these figures, the gas cloud initially comes in from apocenter on a nearly radial orbit.
The cloud's orbital plane is misaligned with the rotation of the background flow, resulting in a perpendicular component to the drag force.
As the cloud approaches the black hole, this perpendicular drag force increases due to both the increasing background density and the increasing rotation velocity.
The resulting torque rapidly re-orients the orbital plane and,
after $\sim$\,5 pericenter passages, the orbital plane is aligned with the rotation axis of the accretion flow.
No further reorientation occurs, but the orbit circularizes and the cloud becomes approximately co-moving with the rotating gas.
The drag force thus drops significantly, and the cloud slowly spirals in toward the black hole.
Of course, after many pericenter passages, the cloud may be tidally distorted to the point where it intersects itself; this process could lead to much more rapid circularization and inflow, but is not accounted for in our calculation.

Figure~\ref{fig:scatter-plots} shows probability distributions of the parameters in our model, assuming the probability defined in equation~\ref{eq:mla-probability} (the standard $\chi^2$ analysis yielded similar results, but with somewhat smaller uncertainties).
The data constrain some parameters much more strongly than others: for example, we localize the orientation of the rotation axis of the accretion flow to within $\sim\,20\deg$.
The rotation axis for the accretion flow we find corresponds to a polar angle $\theta\sim(111\pm16)$° and an azimuthal angle $\phi\sim(-32\pm23)$°.
This is misaligned with the stellar disk by almost 90°, and with G2's orbit by about 60°.
Our analysis does not yet provide an easy identification for the origin of rotation in the galactic center.
Identifying the source of rotation could be further complicated by the back-reaction of the gas cloud onto the accretion flow: since the combined mass of G1 and G2 is comparable to the total mass expected to be in the accretion flow, the drag force on G2's orbit could have rotated the accretion flow by several degrees from its original position.
How the flow responds to this back-reaction depends in detail on momentum transport within the gas; viscosity, convection, turbulence, and magnetic fields could all influence the process significantly.
We have not modeled any such back-reaction in this paper; however it would be straightforward to add once a theory for the background flow is specified.

While our model constrains the rotation axis of the accretion flow, figure~\ref{fig:scatter-plots} shows that other combinations of parameters are degenerate.
The density of the background gas (governed by its power-law slope $\alpha$) and the magnetic field strength (quantified by $\beta$) both affect the magnitude of the drag force.
We can increase the drag force by either raising the density of the gas, so that the cloud accelerates a denser column of ambient gas, or by increasing the field strength, so that magnetic tension couples the cloud to a larger volume of background gas \citep{McCourt2014}.
Since the evolution of the orbit depends only on the overall magnitude of the drag force, we find that the parameters $\alpha$ and $\beta$ are almost completely degenerate.
Similarly, the rotation parameter $f_{\text{kep}}$ and the cloud aspect ratio $L_{\text{cloud}}/R_{\text{cloud}}$ independently control the component of the drag force out of the orbital plane; this perpendicular drag force is proportional to $f_{\text{kep}}^2 \times L_{\text{cloud}}/R_{\text{cloud}}$.
Therefore, we only constrain their product and these parameters are also strongly degenerate.

Thus our model does not directly constrain $\alpha$, $\beta$, $f_{\text{kep}}$, or the shape of the cloud.
However, some of these parameters may be constrained by complementary observations.
The shape of the cloud can be inferred from Br-$\gamma$ observations of the velocity shear; the measurements in \citet{Gillessen2013b} suggest $L_{\text{cloud}}/R_{\text{cloud}}\sim$\,10--20.
This would imply a fairly large rotation rate; $f_{\text{kep}}\sim$\,0.5--1.0.
Similarly, observations of the magnetar in the galactic center \citep{Kennea2013} suggest a magnetic field strength with $\beta\sim{}1$ at large radii \citep[$\sim$\,0.1\,pc;][]{Eatough2013}.
Since $\beta$ is not expected to evolve strongly with radius in many models of the galactic center, this suggests a fairly steep density profile with $\alpha\sim$\,0.7--0.9.
This is within the constraint of the x-ray luminosity, roughly consistent with a radiatively inefficient accretion flow \citep{Yuan2003} and with the density profile inferred from x-ray spectroscopy at larger radii \citep{Wang2013}. 

Interestingly, if we require G2 to be spherical in shape, we cannot reproduce the inferred shift in its orbital plane.
Even with rotation at the Keplerian speed, matching the out-of-plane drag force yields an in-plane force so large that the cloud rapidly spirals into the black hole.
We can only match the measurements if we allow the gas cloud to have a non-spherical shape, so that the drag force points in a direction different than the cloud's relative velocity.
A misalignment between the drag force and the relative velocity is analogous to aerodynamic lift; in a sense, G2 thus partially flies through the accretion flow as it falls toward the black hole.
The drag force we measure implies a glide ratio $\sim{}2$ for G2, comparable to that of the space shuttle ($\sim{}4$).
Our best-fit solutions imply an aspect ratio $L_{\text{cloud}}/R_{\text{cloud}} \sim 10$, similar to what is seen in simulations \citep[\eg][]{Burkert2012,Guillochon2014} and to what is inferred observationally \citep{Gillessen2013b,Pfuhl2015}.
More precise determinations of the shapes of G1 and G2 from fitting their measured velocity shears could thus constrain the rotation parameter $f_{\text{kep}}$; this in turn would provide information about the (unknown) thermodynamics of the accretion flow.

Figure~\ref{fig:kepler-elements} compares the evolution in Kepler elements of well-fitting models as a function of time with derived values for G1 and G2.
The yellow and blue bands indicate the $1$- and $2$-$\sigma$ intervals derived from the analysis shown in figure~\ref{fig:scatter-plots}.
(Within the context of the model defined in equation~\ref{eq:background}, the error bars show the range allowed by the data.  The predictions could of course change by several $\sigma$ under a different evolutionary model.)
The red curves show a random sample of well-fitting solutions -- these yield essentially the same $\chi^2$ and cannot be differentiated using the existing data.
Figure~\ref{fig:kepler-elements} shows that the energy loss is strongly episodic -- the semi-major axis $a$ decreases sharply at pericenter passages, but is nearly constant between them.
The angular momentum increases more continuously, however: $e$ decreases smoothly with time, as do the angles $\Omega'$ and $\omega$.
The inclination $i'$ deceases sharply after the first pericenter passage, then remains constant for several decades.
This provides an alternative test of our model, in addition to the astrometry and velocity curves shown in figure~\ref{fig:2d-orbit}.
Continued observations of $e$ and $\Omega'$ thus represent a promising way to test this model in the coming years and to improve constraints on the parameters for the galactic center accretion flow.

Thus, we have found a model trajectory for gas clouds in the galactic center which matches the observed positions and velocities of G1 and G2, and which is consistent with available constraints on the gas density and magnetic field strength in the galactic center.
This model implies a particular rotation axis for the accretion flow feeding the black hole Sgr~A$^{*}$; in the following section, we describe the orientation of this axis relative to other features in the galactic center.

\section{Discussion}
\label{sec:discussion}
\begin{figure}
  \centering
  \includegraphics[width=\columnwidth]{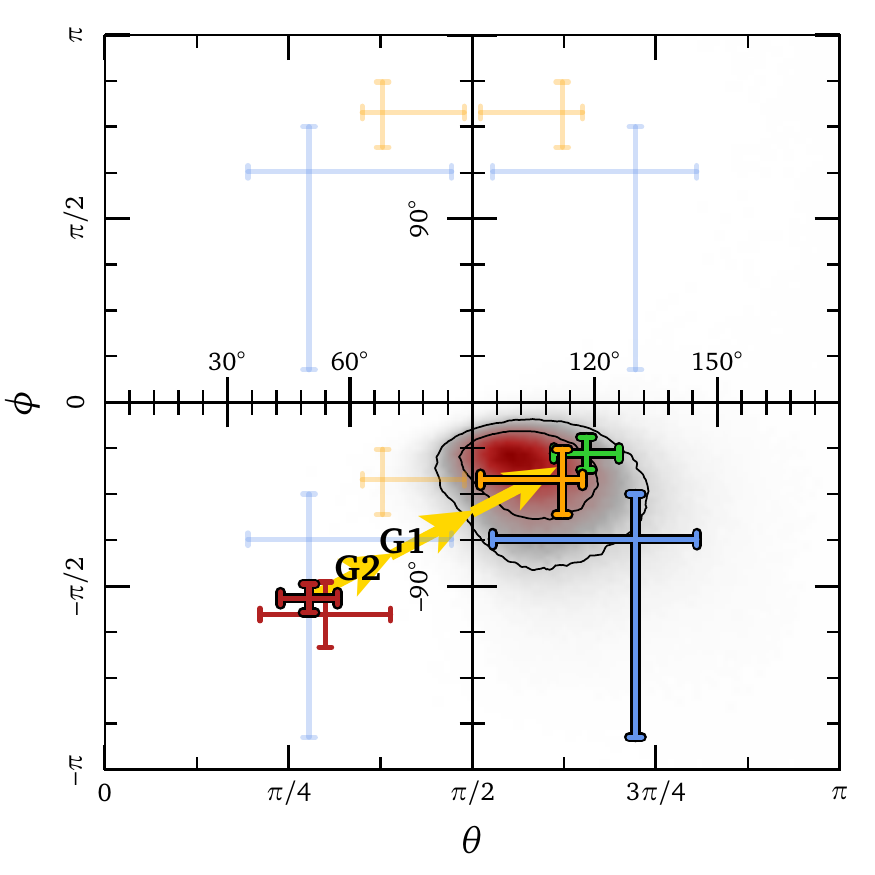}
  \caption{Comparison of our constraint on the rotation axis ({left panel of figure~\ref{fig:scatter-plots}}) with other measurements in the galactic center.
The blue and orange error bars show constraints on the orientation of the horizon-scale accretion disk from measurements by the Event Horizon Telescope (\citealt{Dexter2010} and \citealt{Broderick2011}, respectively).
Currently, the EHT limits the orientation of the rotation axis to one of four possibilities; our constraint is consistent with one of them.  Red error bars mark the orientation and width of the clockwise stellar disk in the galactic center (\citealt{Bartko2009}, large error bars, and \citealt{Yelda2014}, small error bars).
Text markers indicate the orientations of G2 and G1 (table~\ref{tab:compare-fits}).
Though we did not include any information about the clockwise stellar disk in our fit, we find evolution consistent with the G-clouds originating in that disk.
The rotation axis we identify lines up closely with the molecular ring in the galactic center (green error bars; \citealt{Jackson1993}); this supports a scenario in which the accretion flow originates with gas inflowing from the molecular ring.
}\label{fig:theta-phi-annotated}
\end{figure}
The dynamics of the accretion flow around the massive black hole in our galactic center is an important topic of ongoing research.
One fundamental problem in this area is that the Bondi accretion rate implied by x-ray observations at large radii is much higher than the constraints implied by the bolometric luminosity of the gas and by its rotation measure (see \citealt{Yuan2014} for a recent review).
Most of the gas which \textit{could} accrete from large radii therefore never reaches the black hole.
A number of theories have been proposed to explain this deficit, but we can't currently differentiate among them due to a lack of observational constraints on the gas in the galactic center.

\citet{Gillessen2012} reported the discovery of G2, a gas cloud on a highly eccentric orbit about the massive black hole in our galactic center.
G2 has passed through a range of radii where differing models for the galactic center accretion flow yield different predictions for the density and rotation rate of the gas.
Detection of any interaction between G2 and this background gas would therefore provide a critical test of the various theories of galactic center accretion flow.

Following \citet{Pfuhl2015}, we work with the assumption that G2 and the similar cloud G1 follow the same trajectory, with G1 preceding G2 by $\sim$13 years; we quantify in section~\ref{sec:analytic} why this is a reasonable assumption.
Though the orbital parameters of G1 and G2 are strikingly similar, they contain important differences.
Under the assumption that G2 follows G1, these small differences represent evolution due to interaction with the accretion flow, and in turn implies a particular orientation for the drag force acting on the cloud (see table~\ref{tab:jvectors} or figure~\ref{fig:tau-pdf}).
This drag force is inconsistent with motion through a static medium, and thus implies that the accretion flow in the galactic center rotates.
We model this drag force in more detail in section~\ref{sec:numeric}, where we show that we can reproduce the observed trajectories of G1 and G2 with a simple model and reasonable choices for the properties of the galactic center accretion flow (figure~\ref{fig:2d-orbit}).

Figure~\ref{fig:scatter-plots} shows our preliminary results, in which we localize the rotation axis of the accretion flow to within 20°. We show this result in more detail in figure~\ref{fig:theta-phi-annotated} and over-plot earlier constraints on the rotation axis determined by the Event Horizon Telescope \citep[EHT;][]{Dexter2010,Broderick2011,Psaltis2015}.
Degeneracies currently limit the EHT determination to one of four possibilities; our constraint is consistent with only one of them.\footnote{The angle $\phi$ we use in this paper is related to the angle $\xi$, which is measured in degrees east of north, by $\xi = 90 \deg - \phi$.} 
To our knowledge, this paper represents the first unique determination of the rotation axis at intermediate radii ($\sim 10 ~ R_{\text{s}} < R < 10^4 R_s$) in the galactic center.

In figure~\ref{fig:theta-phi-annotated}, we also mark the position of the (inner edge of the) clockwise disk with red error bars \citep{Bartko2009,Lu2009}.
We note that while we did not include information about the clockwise disk in modeling the trajectory of G2, we find evolution entirely consistent with an origin within the clockwise disk.
Yellow arrows sketch the evolution of the clouds in our model: they originate in orbits consistent with the clockwise disk and end up aligned with the rotation axis of the accretion flow.

The rotation axis we identify is in close agreement with the orientation of G359.944--0.052, a linear x-ray filament which points towards Sgr~A$^{*}$.
\citet{Li2013} propose this feature results from a parsec-scale jet emanating from Sgr~A$^{*}$: the jet collides with the Eastern arm of the Sgr~A West H\Rmnum{2} region, driving a shock front which is observable in infrared and radio, and accelerating relativistic electrons which are seen in x-ray.
The long axis of the hypothetical jet path corresponds to an azimuthal angle $\phi\sim{}(-35\pm{}2)$°.
If the jet propagates along the rotation axis, this measurement is consistent with our finding that $\phi\sim(-32\pm23)$°.
The rotation axis is also similar that inferred by jet models of \citet{Markoff2007}, $\theta{}\gtrsim{}75$°, $\phi\sim{}-25$°, and model-dependent constraints on the spin axis by \citet{Meyer2007}, $\phi\sim{}-25$°.
Surprisingly, this rotation axis is also nearly coincident with the galaxy's rotation axis, $\phi=-31.7$° \citep{Reid2004}.

Since we find a rotation axis misaligned with both the stellar clockwise disk (by $\sim{}90$°), and with G2 (by $\sim{}60$°), our model argues against scenarios in which gas in the accretion flow originates from either of these sources.
The rotation axis we identify does align quite closely with the circumnuclear disk, a massive torus of molecular gas with an inner edge $\sim{}1.5$\,pc from Sgr~A$^*$ ($\theta = 118$° $\pm 15$°, $\phi = -25$° $\pm 15$°; \citealt{Jackson1993}).
Our analysis suggests that the gas in the accretion flow hails from the circumnuclear disk, rather than from the young clockwise stellar disk as is commonly assumed.

In addition to constraining the rotation axis of the accretion flow, we find degenerate relationships between the gas density and magnetic field strength, and between the rotation profile and the shape of the G2 cloud.
These degeneracies could be broken with further observations, yielding first measurements of the gas density and rotation speed at radii of 100s of AU in the galactic center.
This is an exciting possibility, because these measurements would constrain the thermodynamics and momentum transport within the accretion flow, providing important information about the physics of black hole accretion in low-luminosity systems such as Sgr~A$^{*}$.

We find that G2's motion is \textit{with} the accretion flow rather than against it \citep[figure~\ref{fig:theta-phi-annotated}; c.f.][]{Sadowski2013a}. With a relative inclination of $\sim$60\deg between G2's pre-pericenter orbit and the accretion flow, this is similar to the M3 simulation of \citet{Abarca2014}. This lowers the relative velocity somewhat, and may help explain non-detections of the bow-shock ahead of the cloud \citep[e.g.][]{Tsuboi2015,Bower2015}. 
A potentially important effect we have not included in our analysis is the back-reaction of the drag force onto the accretion flow; since the combined mass of G1 and G2 is comparable to the total amount of gas in the accretion flow, the drag force acting on the clouds could also influence the evolution of the accretion flow.
This is a complex process, as the back-reaction depends on angular momentum transport in the accretion flow and could be affected by magnetic fields, turbulent transport, and viscosity; we have therefore chosen not to model it in our preliminary study.

Though our model requires the gas cloud to be tidally elongated in its direction of motion, we have not considered the possible disruption of the cloud by tidal forces.
While this disruption is a robust outcome in hydro simulations \citep[e.\,g.][]{Guillochon2014}, G1 evidently survived for at least six years after its pericenter passage.
We therefore assume that some additional process (e.\,g. magnetic tension) enables the clouds to survive pericenter passage.
We will investigate this possibility in the future using magneto-hydrodynamic simulations.

We have not yet considered evolution in the size, shape and mass of the cloud. It is straightforward to generalize our model and let these values evolve with time, given a theoretical or observational prescription. However this is not likely to change our constraints. The change in orbital parameters of the cloud occurs mostly at pericenter where the drag force is strongest. 
Therefore the properties of the cloud far from pericenter have a limited influence on its orbital evolution. Furthermore, the rotation axis of the accretion flow is insensitive to changes in the properties of the cloud. 

Our model is not unique; the data for G1 and G2 can be fit to independent Keplerian orbits, or by the simpler drag model proposed in \citet{Pfuhl2015}.
In both cases, we argue that our model is physically more reasonable.
Crucially, these three models predict very different evolution for G1 in the next $\sim$5--10 years (see figures~\ref{fig:2d-orbit} and~\ref{fig:kepler-elements}).
Continued observations of G1 are therefore essential to test our model and to provide stronger constraints on the properties of the accretion flow.
If our model proves correct, it not only provides a much-needed probe of the gas dynamics at intermediate radii in the galactic center, it also links the dynamics at event horizon-scales to larger radii approaching (and perhaps exceeding) the Bondi radius.

\appendix
\section{Inferring Angular Momentum and Torque Vectors from Observations}
\label{app:fitting}
The angular momentum and torque vectors used in section~\ref{sec:analytic} depend simultaneously on $i$, $\Omega$, $a$, and $e$ for G1 and G2, along with the time delay between their two orbits; quantifying these torques this requires knowledge of the correlated uncertainties in all of these parameters.
We infer the torques directly from the data using an MCMC simulation.
We use the \texttt{emcee} \citep{Foreman2013} implementation of the affine-invariant algorithm presented in \citet{Goodman2010}.
We tried both a standard ``$\chi^2$'' statistic and a maximum-likelihood analysis in which the log-likelihood of a given orbit is proportional to
\begin{align}
  \ln{L} = -\frac{1}{2} \sum_{i} 
  \left[
    \frac{(y_i - x_i)^2}{\sigma_i^2 + (f y_i)^2} 
    + \ln\left[\sigma_i^2  + (f y_i)^2 \right]
  \right],\label{eq:kep-mla}
\end{align}
where the parameter $f$ represents additional uncertainty in the data due to systematic errors; we find similar results in both cases.
Our results are consistent with \citet{Gillessen2013b} and \citet{Pfuhl2015}; table~\ref{tab:compare-fits} and figure~\ref{fig:triangle-g2} show a comparison.

In order to convert from Kepler elements to Cartesian coordinates to compute likelihood functions, we define unit vectors along the eccentricity axis (pointing towards pericenter), along the angular momentum axis (orthogonal to the orbital plane), and the semi-minor axis \citep{Murray1999}.
\begin{subequations}
\begin{align}
  \hat{\vec{e}} &= 
  \left(
  \begin{array}{c}
    \cos(\omega) \cos(\Omega') - \sin(\omega) \cos(i') \sin(\Omega') \\
    \cos(\omega) \sin(\Omega') + \sin(\omega) \cos(i') \cos(\Omega') \\
    \sin(\omega) \sin(i')
  \end{array}
  \right) \\
  \hat{\vec{j}} &= 
  \left(
  \begin{array}{c}
    \sin(i') \sin(\Omega') \\
    -\sin(i') \cos(\Omega') \\
    \cos(i')
  \end{array}
  \right)\\
  \hat{\vec{b}} &= 
  \left(
  \begin{array}{c}
    -\sin(\omega) \cos(\Omega') - \cos(\omega) \cos(i') \sin(\Omega') \\
    -\sin(\omega) \sin(\Omega') + \cos(\omega) \cos(i') \cos(\Omega') \\
    \sin(i') \cos(\omega)
  \end{array}
  \right)
\end{align}
\end{subequations}
Note that this convention differs from that used by, e.\,g. \citet{Gillessen2012} and \citet{Phifer2013}.
We convert between the two coordinate systems with the following transformations:
\begin{subequations}
\begin{align}
  \cos{i'} &= j_{z}/j = -\cos{i} \\
  \tan{\Omega'} 
  &=\frac{(\hat{\vec{z}}\times\vec{j})_y}{(\hat{\vec{z}}\times\vec{j})_x}
  = \cot \Omega
\end{align}
\end{subequations}

We obtain positions and velocities using \citep{Murray1999}:
\begin{subequations}
\begin{align}
\vec{r} &= a (\cos{E} - e) \vec{\hat{e}} + 
           a \sqrt{1 - e^2} \sin{E} \vec{\hat{b}} \\
\vec{v} &= \frac{1}{1 - e \cos{E}}
           \sqrt{\frac{G M}{a}}
           \left(
           \sqrt{1 - e^2} \cos{E} \vec{\hat{b}}
           -e \sin{E} \vec{\hat{e}}
           \right),
\end{align}
\end{subequations}
where the eccentric anomaly $E$ is related to the mean anomaly via $M = E - e \sin{E}$, and the mean anomaly is related to time $t$ via $M = (t - t_{\text{peri}}) \sqrt{{G M}/{a^3}}$.

We do not fit for the uncertainty in the mass or velocity of Sgr~A$^{*}$, nor to the distance to the galactic center; instead, we adopt the best-fit values from \citet{Gillessen2009} and hold these constant throughout our analysis.

\begin{figure}
  \centering
  \includegraphics[width=\columnwidth]{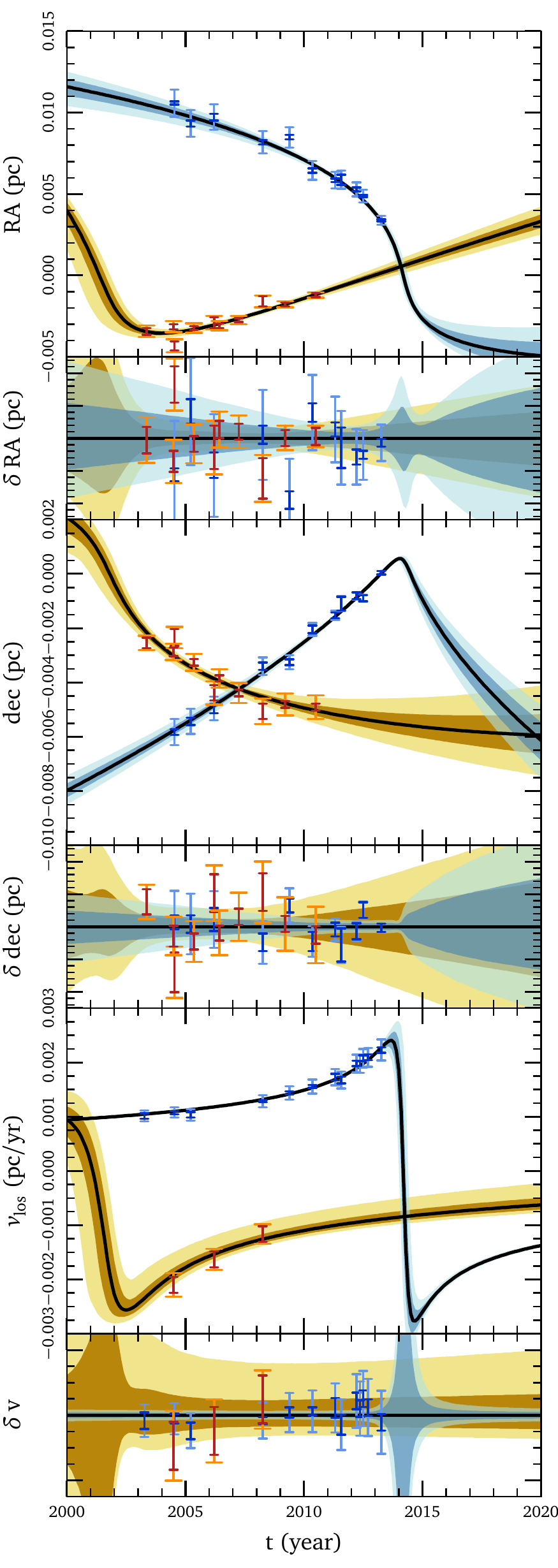}
  \caption{\textit{caption on next column}}\label{fig:fit-uncertainty}
\end{figure}
\begin{figure}
  \contcaption{Keplerian fits to G1 and G2, with comparisons to astrometry and line-of-sight velocity measurements published in \citet{Gillessen2013b} and in \citet{Pfuhl2015}.
From top to bottom, rows show right ascension (RA), declination (dec), and line-of-sight velocity ($v_{\text{los}}$) as functions of time.
Below each plot, we show a residual after subtracting the best-fit model; this enables an easier comparison with the data.
Blue regions show the 1- and 2-$\sigma$ uncertainty in our fit to G2, and yellow curves show the same for G1.
The dark blue points show Br-$\gamma$ astrometry and velocity measurements for G2 published in \citet{Gillessen2013b}.
These are publicly available here: \url{https://wiki.mpe.mpg.de/gascloud/PlotsNData}.
The red points show measurements of G1 published in \citet{Pfuhl2015}; these include both Br-$\gamma$ and L-band astrometry measurements.
Light blue and orange points are the same as the dark blue and red ones, but show the larger error-bars determined in our maximum-likelihood analysis (via the $f$ parameter in equation~\ref{eq:kep-mla}; see section~\ref{subsec:numeric-method}).}
\end{figure}

\begin{figure*}
  \centering
  \includegraphics[height=0.9\textheight]{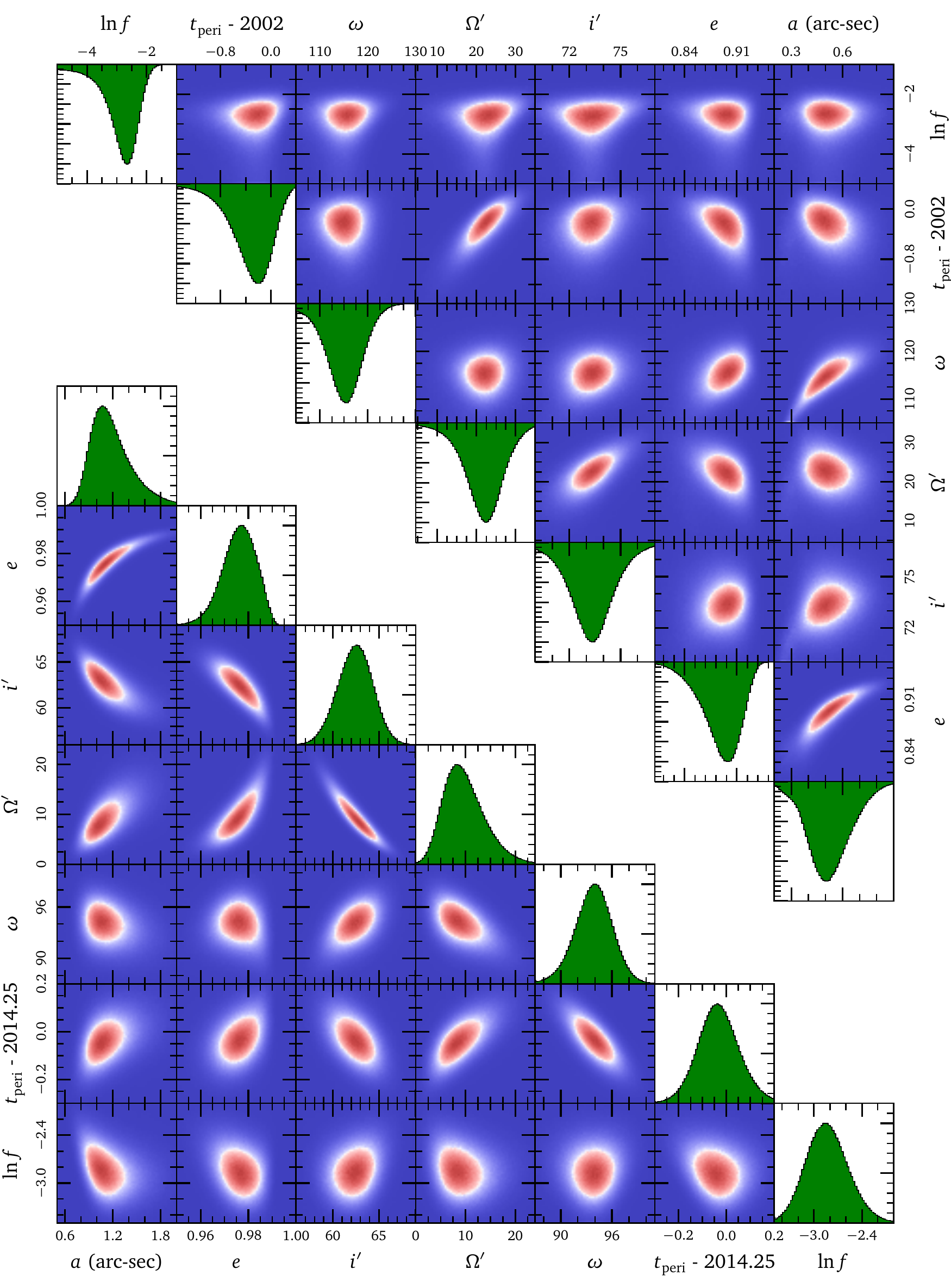}
  \caption{\textit{caption on next page}}\label{fig:triangle-g2}
\end{figure*}
\begin{figure}
  \contcaption{Probability distributions for Keplerian orbital elements for G1 (\textit{top}) and G2 (\textit{bottom}).
Each colored panel shows a projection of the seven-dimensional probability distribution onto the two-dimensional space defined by the indicated coordinates; at the end of each column, the green histogram shows the one-dimensional probability distribution of each variable.
The correlated uncertainties between $i'$ and $\Omega'$ are essential for quantifying the uncertainty in $\hat{\vec{j}}$ and thus $\vec{\tau}_{\perp}$ in section~\ref{sec:analytic}.
Similarly, the correlated uncertainties among all of $a$, $e$, $i'$, and $\Omega'$ are needed to quantify the uncertainty in $\vec{j}$ and $\vec{\tau}$.}
\end{figure}

%
%
%
%

%
%
%
\begin{table*}
\centering
\begin{threeparttable}
\caption{Comparison of our derived Kepler elements with those published in \citet{Pfuhl2015} and \citet{Gillessen2013b}.}\label{tab:compare-fits}
\begin{tabular}
{r@{.}l%
r@{.}l@{\;±\;}r@{.}l
r@{.}l@{\;±\;}r@{.}l%
r@{\;±\;}l%
r@{\;±\;}l%
r@{\;±\;}l%
r@{.}l@{\;±\;}r@{.}l}
\toprule 
%
\multicolumn{2}{c}{year (JD)}
& \multicolumn{4}{c}{$a$\,('')}
& \multicolumn{4}{c}{$e$}
& \multicolumn{2}{c}{$i'$\,(°)}
& \multicolumn{2}{c}{$\Omega'$\,(°)} 
& \multicolumn{2}{c}{$\omega$\,(°)} 
& \multicolumn{4}{c}{$t_{\text{peri}}$} \\
\midrule
%
\multicolumn{2}{c}{G1}
&    0&5  & 0&1   
&    0&89 & 0&03  
&    73   & 1     
&    22   & 5     
&   114   & 4     
& 2001&7  & 0&3   
\\
\multicolumn{2}{c}{G1\tnote{1}}
&    0&4  & 0&2   
&    0&86 & 0&05  
&    72   & 2     
&    21   & 5     
&   109   & 8     
& 2001&6  & 0&4   
\\
\midrule
%
%
%
\multicolumn{2}{c}{G2}
&    1&2   & 0&2    
&    0&976 & 0&007  
&    62    & 2      
&     9    & 4      
&    94    & 2      
& 2014&22  & 0&08   
\\
\multicolumn{2}{c}{G2\tnote{2}}
&    1&0   & 0&2    
&    0&976 & 0&007  
&    62    & 2      
&     8    & 4      
&    97    & 2      
& 2014&25  & 0&06   
\\
\bottomrule
\end{tabular}
\begin{tablenotes}
\item [1] Pfuhl et al. 2014
\item [2] Gillessen et al. 2013b
\end{tablenotes}
\end{threeparttable}
\end{table*}


\section*{Acknowledgments}

\noindent{}We thank James Guillochon, Ramesh Narayan, and Eliot Quataert for interesting and helpful conversations which shaped the direction of this work.
Michael Johnson provided helpful comments, particularly relating to the EHT orientation measurement.
A.-M.M. is supported by the National Aeronautics and Space Administration through Einstein Postdoctoral Fellowship Award Number PF2-130095 issued by the Chandra x-ray Observatory Center, which is operated by the Smithsonian Astrophysical Observatory for and on behalf of the National Aeronautics Space Administration under contract NAS8-03060.
M.M. is supported by the National Science Foundation grant AST-1312651.
The authors acknowledge the Texas Advanced Computing Center (TACC) at The University of Texas at Austin for providing HPC resources that have contributed to the research results reported within this paper.
This work used the Extreme Science and Engineering Discovery Environment (XSEDE allocations TG-AST140039, TG-AST140047, and TG-AST140083), which is supported by National Science Foundation grant number ACI-1053575.

\bibliographystyle{mn2e}
\bibliography{drag-force}

\label{lastpage}
\end{document}